%% file: bayesian.tex
\documentclass[10pt]{iopart}
\input{packages}

\begin{document}
\title{Bayesian detection of unmodeled bursts of gravitational waves.}

\date{\today}

\author{Antony C. Searle$^1$, Patrick J. Sutton$^2$, Massimo Tinto$^3$}


\address{$^1$LIGO - California Institute of Technology, Pasadena, CA 91125}
\address{$^2$School of Physics and Astronomy, Cardiff University, Cardiff CF24 3AA, United Kingdom}
\address{$^3$Jet Propulsion Laboratory, California Institute of Technology, Pasadena, CA 91109}

\ead{acsearle@ligo.caltech.edu}

\begin{abstract}
  The data analysis problem of coherently searching for unmodeled
  gravitational-wave bursts in the data generated by a global network
  of gravitational-wave observatories has been at the center of
  research for almost two decades. As data from these detectors is
  starting to be analyzed, a renewed interest in this problem has been
  sparked.  A Bayesian approach to
  the problem of coherently searching for gravitational wave bursts with
  a network of ground-based interferometers is here presented.
  We demonstrate how to systematically incorporate prior information
  on the burst signal and its source into the analysis.  This information
  may range from the very minimal, such as best-guess durations,
  bandwidths, or polarization content, to complete prior knowledge of the
  signal waveforms and the distribution of sources through spacetime.
  We show that this comprehensive Bayesian formulation contains
  several previously proposed detection statistics as special
  limiting cases, and demonstrate that it outperforms them.
\end{abstract}




\maketitle

\input{introduction}

\input{analysis}

\input{results}

\input{conclusions}

\section{Acknowledgments}
  We would like to thank Shourov Chatterji, Albert Lazzarini, Soumya Mohanty,
  Andrew Moylan, Malik Rakhmanov, and Graham Woan for useful discussions and
  valuable comments on the manuscript.  This work was performed under partial
  funding from the following NSF Grants: PHY-0107417, 0140369,
  0239735, 0244902, 0300609, and INT-0138459.  A.~Searle was supported
  by the Australian Research Council and the LIGO Visitors Program.
  For M.~Tinto, the research was also performed at the Jet Propulsion
  Laboratory, California Institute of Technology, under contract with
  the National Aeronautics and Space Administration.  M.~Tinto was
  supported under research task 05-BEFS05-0014.  P.~Sutton was
  supported in part by STFC grant PP/F001096/1.  LIGO was constructed
  by the California Institute of Technology and Massachusetts
  Institute of Technology with funding from the National Science
  Foundation and operates under cooperative agreement PHY-0107417.
  This document has been assigned LIGO Laboratory document number
  LIGO-P070114-00-Z.

\section{References}
\bibliography{master}
\bibliographystyle{unsrt}
\end{document}

%% file: packages.tex
\usepackage{amssymb}
\usepackage{graphicx}
\graphicspath{{figures/}}
\usepackage{subfigure}
\usepackage[usenames]{color}
\usepackage[normalem]{ulem}

%% file: introduction.tex
\section{Introduction}
\label{sec:introduction}

Large-scale, broad-band interferometric gravitational-wave
observatories \cite{GEO,LIGO,TAMA,VIRGO} are operating at their
anticipated sensitivities, and scientists around the globe have begun
to analyze the data generated by these instruments in search of
gravitational wave signals \cite{Ab_etal:07rh}.  Gravitational wave
bursts (GWBs) are among the most exciting signals scientists expect to
observe, as our present knowledge and modeling ability of GWB-emitting
systems is rather limited. These signals often depend on complicated
(and interesting) physics, such as dynamical gravity and the equation
of state of matter at nuclear densities.  While this makes GWBs an
especially attractive target to study, our lack of knowledge also
limits the sensitivity of searches for GWBs. Potential sources of GWBs
include merging compact objects
\cite{FlHu:98a,FlHu:98b,Pr:05,Ba_etal:06, Ca_etal:06,Di_etal:06,
  HeShLa:06,LoReAn:06,ShTa:06}, core-collapse supernovae
\cite{ZwMu:97,DiFoMu:02b,OtBuLiWa:04,ShSe:04,OtBuDeLi:06}, and
gamma-ray burst engines \cite{Me:02}; see \cite{CuTh:02} for an
overview.

Although a gravitational wave signal is characterized by two
independent polarizations, an interferometric gravitational wave
detector is sensitive to a single linear combination of them.
Simultaneous observations of a gravitational wave burst by three or
more observatories over-determines the waveform, permitting the
source position on the sky to be determined, and a least-squares
estimate of the two polarizations.  This solution of the {\em inverse
problem} for gravitational waves was first derived by G\"{u}rsel and
Tinto \cite{GuTi:89} for three interferometers.  Subsequent work
generalized it \cite{Tinto96} and  formalized it as an example of
a coherent maximum likelihood statistic by Flanagan and Hughes \cite{FlHu:98b} and later Anderson, Brady,
Creighton, and Flanagan \cite{AnBrCrFl:01}.
Various modifications have been proposed, such as Rakhmanov's Tikhonov
\cite{Ra:06} and Summerscale's maximum-entropy \cite{Summerscales:2007xq} regularization
techniques, Kilmenko \emph{et al}'s constraint likelihood method \cite{KlMoRaMi:05,MoRaKlMi:06}, and the SNR variability approach of
Mohanty {\em et al.} \cite{MoRaKlMi:06}.  Potential as a consistency test
was noted by Wen and Schutz in \cite{WeSc:05} and demonstrated in
Chatterji \emph{et al} \cite{Ch_etal:06}.  Other
coherent detection algorithms have been proposed by Sylvestre
\cite{Sylvestre:03} and by Arnaud {\em et al.} \cite{Arnaud:03}.

These approaches to signal detection have generally been derived
by following either \emph{ad hoc} reasoning or a maximum-likelihood
criterion (notable exceptions, foreshadowing our Bayesian approach, are Finn \cite{finn:97},
Anderson \emph{et al} \cite{AnBrCrFl:01} and Allen \emph{et al} \cite{Al_etal:03}).
In this paper we present a systematic and comprehensive
Bayesian formulation \cite{jaynes, gregory}
of the problem of coherent detection of gravitational wave
bursts.  We demonstrate how to incorporate partial or incomplete
knowledge of the signal in the analysis, thereby improve the probability
of detection of these weak signals.  This information may include
time-frequency properties of the signal, polarization content,
model waveform families or templates, as well as information on the
distribution of the source through spacetime.  We also explicitly identify
the prior assumptions that must be made about the signal to cause a
Bayesian analysis to behave like several of the previously proposed
detection statistics.

Real interferometers also experience instrumental artifacts that can masquerade
as signals.  These are typically dealt with in post-processing rather than by
the detection statistic itself, but some proposals have been made to include a
model for these ``glitches'' in the detection statistic
\cite{PrPi:08,SeSuTiWo:08}.  An advantage of the Bayesian framework is that the standard
choice between signal and stationary noise could be extended to include a third option: randomly occurring noise ``glitches''.  This
formulation is a promising direction for future progress, but we do not further
discuss it in this paper.

The paper is organized as follows.
In \S\ref{sec:analysis} we derive the Bayesian posterior detection
probability of an idealized delta-like burst signal by a
toy-model network of observatories.
Greatly expanding on \cite{SeSuTiWo:08}, we then generalize the derivation of the
Bayesian odds ratio into a usable statistic for an arbitrary number of
interferometers with differently colored and (potentially) correlated
noises. We also consider a wide range of signal models corresponding
to different states of knowledge about the burst, from total
ignorance to complete {\em a priori} knowledge of the waveforms.
In \S\ref{sec:simulations} we characterize the relative performance of
previously proposed statistics (the G\"ursel-Tinto/standard and constraint likelihoods)
and the Bayesian statistic by
performing a Monte-Carlo simulation in which we add a simple binary
black-hole merger waveform to simulated detector data and construct
Receiver-Operating Characteristic (ROC) curves. We find that the
Bayesian method increases the probability of detection
for a given false alarm rate by approximately 50\%, over those
associated with previously proposed statistics.

%% file: analysis.tex
\section{Analysis}
\label{SECII}\label{sec:analysis}


\subsection{Single-sample observation}
\label{sec:singleSample}

We begin by investigating perhaps the simplest Bayesian coherent data
analysis: detecting a signal from a known sky position in a single
strain sample from each of $N$ gravitational wave observatories.  This
example will show many of the basic features of the Bayesian analysis,
and highlight some of the differences between the Bayesian approach
and previous statistics.  In the following section we will generalize
to a multi-sample search for a signal arriving at an unknown time from
an unknown sky position.

Consider a single strain sample from each of $N$ detectors, each
measurement taken at the moment corresponding to the passage of a
postulated plane gravitational wave from some known location on
the sky, ($\theta, \phi$).  The measurements are then equal to \cite{GuTi:89}
\begin{equation}
\mathbf{x}=\mathbf{F}\,\mathbf{h}+\mathbf{e} \, , \label{eqn:ssmodel}
\end{equation}
where $\mathbf{x}$ is the vector of measurements $[x_1,\ldots,x_N]^T$, the
matrix $\mathbf{F}=[[F_1^+,F_1^\times],\ldots,[F_N^+,F_N^\times]]$
contains the antenna responses of the observatories to the postulated
gravitational wave strain vector $\mathbf{h}=[h_+,h_\times]^T$, and
$\mathbf{e}$ is the noise in each sample.  $\mathbf{F}$ is a known
function of the source sky direction $(\theta,\phi)$, and the decomposition into $+$ and $\times$ polarizations requires us to choose an arbitrary polarization basis angle $\psi$ for each source sky direction.

We wish to distinguish between two hypotheses: $H_0$,
that the data contains only noise, and $H_1$, that the
data contains a gravitational wave signal.  The Bayesian odds ratio \cite{jaynes, gregory}
allows us to compare the plausibility of the hypotheses:
\begin{equation}
\frac{p(H_1|\mathbf{x},I)}
{p(H_0|\mathbf{x},I)}=
\frac{p(H_1|I)}
{p(H_0|I)}
\frac{p(\mathbf{x}|H_1,I)}
{p(\mathbf{x}|H_0,I)}
\label{Bayes_Ratio} \, ,
\end{equation}
where $I$ is a set of unstated but shared assumptions (such as the
detector locations, orientations and noise power spectra). If the posterior plausibility ratio is greater than one,
$H_1$ is more plausible than $H_0$ and we
classify the observation as a detection.  If the posterior
plausibility ratio is less than one, $H_1$ is less
plausible than $H_0$ and we classify the observation as a
non-detection.

The $p(H|I)$ terms (``plausibility of $H$ assuming $I$'')are the
\emph{prior} plausibilities we assign to each hypothesis $H$ on the
basis of our knowledge $I$ prior to considering the measurement; for
example, our expectation that detectable gravitational waves are rare
requires that $p(H_1|I)\ll p(H_0|I)$.

The $p(\mathbf{x}|H,I)$ terms (``plausibility of $\mathbf{x}$ assuming
$H$ and $I$'') are the probabilities assigned by a hypothesis to the
occurrence of a particular observation $\mathbf{x}$.  These are
sometimes called likelihood functions; they represent the likelihood
of a certain measurement being made.

The $p(H|\textbf{x},I)$ terms are the \emph{posterior} plausibilities
we assign to the hypotheses in light of the observation.
The hypothesis that assigned more probability to the observation becomes more plausible.

For notational simplicity we will drop the $I$ in our formulae; the unstated assumptions are implicit.

If we make the idealized assumption that the noise in each detector is
independent and normally distributed \cite{jaynes, gregory} with zero mean and unit standard
deviation, we can then write the following expression for the
likelihood $p(\mathbf{x}|H_0)$
\begin{eqnarray}
p(\mathbf{x}|H_0)&=&\prod_{i=1}^N p(x_i|H_0)\nonumber\\
&=&\prod_{i=1}^N\frac{1}{\sqrt{2\pi}}\exp(-\frac{1}{2}x_i^2)\nonumber\\
&=&(2\pi)^{-\frac{N}{2}}\exp(-\frac{1}{2}\mathbf{x}^T\mathbf{x})\label{singleNoise} \, ,
\label{noise_only}
\end{eqnarray}
where $^T$ denotes matrix transposition.  For real detectors, the
measurements can be \emph{whitened}, which modifies the effective beam pattern functions
$\mathbf{F}$.

If we assume that there is a gravitational wave $\mathbf{h}$ present, then
after subtracting away the response $\mathbf{F}\,\mathbf{h}$ the data will
be distributed as noise and the likelihood
$p(\mathbf{x}|\mathbf{h},H_1)$ becomes
%
%
\begin{eqnarray}
p(\mathbf{x}|\mathbf{h},H_1)
&=&(2\pi)^{-\frac{N}{2}}\exp(-\frac{1}{2}(\mathbf{x}-\mathbf{F}\,\mathbf{h})^T
(\mathbf{x}-\mathbf{F}\,\mathbf{h})) \label{noiseSignal} \, .
\label{noise_signal}
\end{eqnarray}

Unfortunately, we do not know the signal strain vector $\mathbf{h}$
{\em a priori}.  To compute the plausibility of the more general
hypothesis $p(\mathbf{x}|H_\mathrm{signal})$ we need to marginalize
away these {\it nuisance parameters}
\begin{eqnarray}
p(\mathbf{x}|H_1)
&=&\int_{-\infty}^{+\infty} \int_{-\infty}^{+\infty} p(\mathbf{h}|H_1)
p(\mathbf{x}|\mathbf{h},H_1) \, \mathrm{d}{h_+} \, \mathrm{d}{h_\times} \, .
\label{marginal}
\end{eqnarray}
The hypothesis resulting from the marginalization integral is an
average of the hypotheses for particular signals $\mathbf{h}$,
weighted by the prior probability $p(\mathbf{h}|H_\mathrm{signal})$ we assign
to those signals occurring.  A convenient choice of prior is to use a normal
distribution for each polarization, with a standard deviation $\sigma$
indicative of the amplitude scale of gravitational waves we hope to
detect. Under these assumptions the prior is
\begin{eqnarray}\label{wave_distribution}
p(\mathbf{h}|H_1)
& = &
        \frac{1}{2\pi\sigma^2}\exp(-\frac{1}{2\sigma^2}\mathbf{h}^T\mathbf{h}) \, .
\end{eqnarray}
This allows us to perform the marginalization integral analytically
\begin{eqnarray}
p(\mathbf{x}|H_1)
& = &
        (2\pi)^{-\frac{N}{2}-1}\sigma^{-2} \int_{-\infty}^{+\infty} \int_{-\infty}^{+\infty}
       \exp(-\frac{1}{2}((\mathbf{x}-\mathbf{F}\,\mathbf{h})^T
        (\mathbf{x}-\mathbf{F}\,\mathbf{h})
        \nonumber \\
&   &   \mbox{}
        +\sigma^{-2}\mathbf{h}^T\mathbf{h})) \, \mathrm{d}{h_+} \, \mathrm{d}{h_\times}
        \nonumber \\
& = &
        (2\pi)^{-\frac{N}{2}}
        |\mathbf{I-K_\mathrm{ss}}|^{\frac{1}{2}} \exp(-\frac{1}{2}\,\mathbf{x}^T
        (\mathbf{I-K_\mathrm{ss}})\mathbf{x}) \, ,
        \label{eq:simpleP}
\end{eqnarray}
%
%
where
\begin{eqnarray}
\mathbf{K_\mathrm{ss}}
&\equiv&
        \mathbf{F}
        (\mathbf{F}^T\mathbf{F}+\sigma^{-2}\mathbf{I})^{-1}
        \mathbf{F}^T\label{eq:simpleC}.
\end{eqnarray}
The result is a multivariate normal distribution with covariance
matrix $(\mathbf{I-K_\mathrm{ss}})^{-1}$, which quantifies the correlations among the
detectors due to the presence of a gravitational wave signal.

With both hypotheses defined, we can form the \emph{likelihood ratio}
\begin{eqnarray}
\Lambda
& = &
        \frac{p(\mathbf{x}|H_1)}
        {p(\mathbf{x}|H_0)}
        \nonumber\\
& = &
        |\mathbf{I-K_\mathrm{ss}}|^\frac12 \exp ( \frac{1}{2}\,\mathbf{x}^T
        \mathbf{F}(\mathbf{F}^T\mathbf{F}+\sigma^{-2}\mathbf{I})^{-1}
        \mathbf{F}^T\mathbf{x}) \, . \,
\label{eqn:ssLambda}
\label{likelihood_final}
\end{eqnarray}
Multiplying the likelihood ratio by the prior plausibility ratio
$p(H_1)/p(H_0)$ completes the calculation of the Bayesian odds ratio
(\ref{Bayes_Ratio}).

In the limit
$\sigma\rightarrow\infty$ we find that the odds ratio contains the
least-squares estimate of the strain
\begin{eqnarray}
\mathbf{\hat{h}}&=&(\mathbf{F}^T\mathbf{F})^{-1}\mathbf{F}^T\mathbf{x} \, .
 \end{eqnarray}
The odds ratio may then be rewritten in terms of a matched filter for the
response to the estimated strain, $\mathbf{x}^T\mathbf{F}\,\mathbf{\hat{h}}$.
For finite values of $\sigma$, the odds ratio contains the \emph{Tikhonov regularized}
estimate of the strain \cite{Ra:06}
\begin{eqnarray}
  \mathbf{\hat{h}} = (\mathbf{F}^T\mathbf{F}+\sigma^{-2}\mathbf{I})^{-1}\mathbf{F}^T\mathbf{x} \, ,
\end{eqnarray}
and can still be rewritten as a matched filter for this estimate.

It is also worth noting the presence in (\ref{eqn:ssLambda}) of the determinant $|\mathbf{I-K_\mathrm{ss}}|$ factor.
It is independent of the data and depends only on the antenna pattern and the signal model.  In particular, it tells us how strongly
to weight likelihoods computed for different possible sky positions
of the signal.  This {\em Occam factor} penalizes sky positions
of high sensitivity relative to sky positions of lower sensitivity which
give similar exponential part of the likelihood.  The effect is typically small compared to the
exponential in most cases if the data has good evidence for a signal,
but can be important for weak signals and for parameter estimation.


\subsection{General Bayesian model}

We now generalize the analysis of the previous section to the case of
burst signals of extended duration and unknown source sky direction $(\theta, \phi)$ and arrival
time $\tau$ with respect to the centre of the Earth.

A global network of $N$ gravitational wave detectors each produce a
time-series of $M$ observations with sampling frequency
$f_\textrm{s}$, which we pack into a single vector
\begin{equation}
\fl
\mathbf{x}=[x_{1,1},x_{1,2},\ldots,x_{1,M},x_{2,1},x_{2,2},\ldots,x_{2,M},\ldots,x_{N,1},x_{N,2},\ldots,x_{N,M}]^T \ .
\end{equation}
Our signal model is a generalization of (\ref{eqn:ssmodel}),
\begin{eqnarray}
\mathbf{x}&=&\mathbf{F}(\tau,\theta,\phi)\cdot\mathbf{h}+\mathbf{e} \, ,\label{eq:linearmodel}
\end{eqnarray}
where
\begin{eqnarray}
\mathbf{h}&=&[h_{+,1},h_{+,2},\ldots,h_{+,L},h_{\times,1},\ldots,h_{\times,L}]^T
\end{eqnarray}
is a time-series of $2 L$ samples describing the band-limited strain
waveform (with the two polarizations packed into a single vector),
$\mathbf{e}$ is a random variable representing the
instrumental noise, and $\mathbf{F}(\tau,\theta,\phi)$ is a $NM\times
2L$ response matrix describing the response of each observatory to an
incoming gravitational wave,
\begin{eqnarray}
\fl \mathbf{F}(\tau,\theta,\phi)&=&
\left[
\begin{array}{cc}
F^+_1(\theta,\phi)\mathbf{T}(\tau+\Delta\tau_1(\theta,\phi)) & F^\times_1(\theta,\phi)\mathbf{T}(\tau+\Delta\tau_1(\theta,\phi)) \\
F^+_2(\theta,\phi)\mathbf{T}(\tau+\Delta\tau_2(\theta,\phi)) & F^\times_2(\theta,\phi)\mathbf{T}(\tau+\Delta\tau_2(\theta,\phi)) \\
\vdots & \vdots \\
F^+_N(\theta,\phi)\mathbf{T}(\tau+\Delta\tau_N(\theta,\phi)) & F^\times_N(\theta,\phi)\mathbf{T}(\tau+\Delta\tau_N(\theta,\phi))
\end{array}
\right] \, .
\end{eqnarray}
Each $M\times L$ block of the response matrix is responsible for
scaling and time shifting one of the waveform polarizations for one
detector, so each block is the product of the directional
sensitivity of each detector to each polarization, $F^+_i(\theta,\phi)$
or $F^\times_i(\theta,\phi)$, and a time delay matrix $T_{j,k}(t)$
\footnote{
From the assumption that the signal is band-limited, it follows that the
time delay matrix may be written as $T_{j,k}(t)=\textrm{sinc}(\pi(j-k-f_\textrm{s}t))$; for $L = M$ and zero time delays, it is equal to the identity matrix; for $L = M$ and time delays corresponding
to integer numbers of time samples, it is a \emph{shift matrix}.
},
 for the source sky direction
dependent arrival times $\tau+\Delta\tau_i(\theta,\phi)$ at each
detector.

\subsection{Noise model}

The noise that affects gravitational wave detectors is typically
modeled as stationary, colored gaussian noise that is independent of the signal parameters.  This can be represented with a
\emph{multivariate normal distribution}, which can be compactly written as
\begin{eqnarray}
\mathcal{N}(\mathbf{\mu},\mathbf{\Sigma},\mathbf{x})&=&\frac{1}{(2\pi)^{N/2}\sqrt{|\mathbf{\Sigma}|}}\exp(-\frac{1}{2}(\mathbf{x}-\mathbf{\mu})^T\mathbf{\Sigma}^{-1}(\mathbf{x}-\mathbf{\mu})) \, .
\end{eqnarray}
The vector $\mathbf{\mu}$ is the mean of the distribution, and the
positive-definite \emph{covariance matrix} $\mathbf{\Sigma}$
describes the ellipsoidal shape of the constant-density contours of the distribution in terms of the
pairwise covariances of the samples,
\begin{eqnarray}
\mathbf{\Sigma}_{i,j} = \langle(e_i-\mu_i),(e_j-\mu_j)\rangle \, .
\end{eqnarray}
Using this notation, the noise likelihood is 
\begin{eqnarray}
p(\mathbf{x}|H_0) &=& \mathcal{N}(\mathbf{0},\mathbf{\Sigma},\mathbf{e})
\end{eqnarray}
for some $MN\times MN$ positive definite matrix $\mathbf{\Sigma}$.
Under the additional assumption of stationarity over some timescale,
these covariances can be estimated from previous observations.

In the case of Gaussian stationary colored noise, each detector is individually
represented by a Toeplitz covariance matrix $\mathbf{\Sigma}^{(i)}$.  For uncorrelated noise,
the covariance matrix for the whole network is $\mathbf{\Sigma} =
\textrm{diag}(\mathbf{\Sigma}^{(1)},
\mathbf{\Sigma}^{(2)},\ldots,\mathbf{\Sigma}^{(N)})$.  In the simple
case in which all the noises are white, have equal standard deviation and are uncorrelated, we have $\mathbf{\Sigma} =
\textrm{diag}(\mathbf{I}, \mathbf{I},\ldots,\mathbf{I})=\mathbf{I}$.

The generalization of (\ref{noise_signal}) and (\ref{marginal}) for the signal likelihood is
\begin{eqnarray}
p(\mathbf{x}|H_1)
&=&
\int_{V_{\mathbf{h},\tau,\theta,\phi}} \!\!\!\!\!\!\!\!\!
\mathcal{N}(\mathbf{F}(\tau,\theta,\phi)\cdot\mathbf{h},\mathbf{\Sigma},\mathbf{x}) \,
p(\mathbf{h},\tau,\theta,\phi|H_1) \,
\mathrm{d}\mathbf{h}\ldots\mathrm{d}\phi \ ,\label{eq:partialmarginalization}
\end{eqnarray}
where ${V_{\mathbf{h},\tau,\theta,\phi}}$ is the space of all signal parameters
and $p(\mathbf{h},\tau,\theta,\phi|H_1)$ is the prior for these parameters.
Without loss of generality we may separate this signal prior into a
prior on source sky direction and arrival time, and a prior on the waveform
\emph{conditional on} the source sky direction and the arrival time, i.e.
\begin{eqnarray}
p(\mathbf{h},\tau,\theta,\phi|H_1)
 = p(\tau,\theta,\phi|H_1) \, p(\mathbf{h}|\tau,\theta,\phi,H_1) \, ,
\end{eqnarray}
giving
\begin{eqnarray}
\fl p(\mathbf{x}|H_1)
&=&
\int_{V_{\mathbf{h},\tau,\theta,\phi}} \!\!\!\!\!\!\!\!\!
\mathcal{N}(\mathbf{F}(\tau,\theta,\phi)\cdot\mathbf{h},\mathbf{\Sigma},\mathbf{x}) \,
p(\tau,\theta,\phi|H_1) \, p(\mathbf{h}|\tau,\theta,\phi,H_1) \,
\mathrm{d}\mathbf{h}\ldots\mathrm{d}\phi \ .\label{eq:partialmarginalization2}
\end{eqnarray}

\subsection{Wideband signal model}
\label{sec:wideband}

In analogy with the single sample case, we can choose a multivariate normal distribution prior for the waveform amplitudes and
render the integral soluble in closed form.
The marginalization integral over $\mathbf{h}$ in (\ref{eq:partialmarginalization2}) can then be analytically performed, giving
\begin{eqnarray}
\frac{p(\mathbf{x}|\tau,\theta,\phi,H_1)}
{p(\mathbf{x}|H_0)}
&=&
\frac{
\int_{\mathbb{R}^{2L}}
\mathcal{N}(\mathbf{F}(\tau,\theta,\phi)\cdot\mathbf{h},\mathbf{\Sigma},\mathbf{x}) \,
p(\mathbf{h}|\tau,\theta,\phi,H_1) \,
\mathrm{d}\mathbf{h}}
{
\mathcal{N}(\mathbf{0},\mathbf{\Sigma},\mathbf{x})
} \label{eq:quick}
\end{eqnarray}
(see (\ref{eqn:explicit}) below).  Numerical integration over a more
manageable three dimensions is then sufficient to compute the Bayes factor,
\begin{eqnarray}
\frac{
p(\mathbf{x}|H_1)
}{
p(\mathbf{x}|H_0)
}
&=&
\int\int\int p(\tau,\theta,\phi|H_1) \,
\frac{p(\mathbf{x}|\tau,\theta,\phi,H_1)}
{p(\mathbf{x}|H_0)} \,
\mathrm{d}\tau \, \mathrm{d}\theta \, \mathrm{d}\phi \, .
\end{eqnarray}
This signal model is computationally tractable.  It represents signals that can be described by an invertible $2 L\times 2 L$ correlation matrix,
including the important 'least informative' case of independent, normally distributed samples of $\mathbf{h}$.


\subsection{Informative signal models}


The wideband signal model excludes some important cases, such as when we have a known waveform, almost known waveform
(such as from a family of numerical simulations) or even just a signal restricted to some frequency-band.  These signals
are superpositions of a (relatively) small number $G < 2 L$ of basis waveforms, that may themselves be characterized
by a finite number of parameters, which we denote $\rho$.
These parameters must be numerically integrated,
like $\tau$, $\theta$, and $\phi$, which may be time-consuming.
Their prior distribution will be denoted by
$p(\mathbf{\rho}|\tau,\theta,\phi,H_1)$.

To describe the signal as a superposition of basis waveforms \cite{Heng:09},
define a set of amplitude parameters $\mathbf{a}$ mapped into strain
$\mathbf{h}$ via a $2L\times G$ matrix
$\mathbf{W}(\rho,\tau,\theta,\phi)$ whose columns
$\mathbf{w}_i(\rho,\tau,\theta,\phi)$ are the basis waveforms, so that
\begin{eqnarray}
\mathbf{h}&=&\mathbf{W}(\mathbf{\rho},\tau,\theta,\phi)\cdot\mathbf{a} \ .
\end{eqnarray}
We assume that the amplitude parameters $\mathbf{a}$ are multivariate normal distributed with a
covariance matrix $\mathbf{A}(\mathbf{\rho},\tau,\theta,\phi)$, so that
\begin{eqnarray}
p(\mathbf{a}|\mathbf{\rho},\tau,\theta,\phi,H_1)&=&\mathcal{N}(\mathbf{0},\mathbf{A}(\mathbf{\rho},\tau,\theta,\phi),\mathbf{a}) \, .
\end{eqnarray}
The resulting distribution for the waveform strain is
\begin{eqnarray}
\fl p(\mathbf{h}|\tau,\theta,\phi,H_1)
&=&
\int_{V_{\rho}} \int_{\mathbb{R}^{G}}
p(\mathbf{h}|\mathbf{a},\mathbf{\rho},\tau,\theta,\phi,H_1) \,
p(\mathbf{a},\mathbf{\rho}|\tau,\theta,\phi,H_1)
\, \mathrm{d} \mathbf{a} \, \mathrm{d} \mathbf{\rho} \, \nonumber \\
 &=&
\int_{V_{\rho}} \int_{\mathbb{R}^{G}}
\delta(\mathbf{h}-\mathbf{W}\cdot\mathbf{a}) \,
\mathcal{N}(\mathbf{0},\mathbf{A},\mathbf{a}) \,
p(\mathbf{\rho}|\tau,\theta,\phi,H_1)
\, \mathrm{d}\mathbf{a} \, \mathrm{d}\mathbf{\rho} \, ,
\end{eqnarray}
where for clarity we have begun to omit the dependence of matrices on their
parameters.  As $G < 2L$ ({\em i.e.}, we have fewer basis waveforms than
samples in the signal time-series) the integral over $\mathbf{a}$ cannot
be directly represented as a multivariate normal distribution.

This signal model proposes that gravitational wave signals have
waveforms that are the sum of $G$ basis waveforms with amplitudes that
are normally distributed (and potentially correlated).  The basis
waveforms and their amplitude distributions may vary with source sky direction,
arrival time, and any other parameters we care to include in
$\mathbf{\rho}$.  The model is capable of representing a variety of
sources including the important special cases of known `template'
waveforms, and band-limited bursts.  We will consider some
concrete examples in \S\ref{sec:signalexamples}; perhaps the most
important is a scale parameter $\sigma$, that permits us to look
for signals of different total energies.

We can substitute the expression back into part of
(\ref{eq:quick}) to form a multivariate normal distribution
partial integral whose solution is given in \cite{jaynes}:
\begin{eqnarray}
\fl p(\mathbf{x}|\tau,\theta,\phi,H_1)
&=&
\int_{V_{\rho}}\int_{\mathbb{R}^{G+2L}} \!\!\!\!
\mathcal{N}(\mathbf{F}\cdot\mathbf{h},\mathbf{\Sigma},\mathbf{x}) \,
\delta(\mathbf{h}
-\mathbf{W}\cdot\mathbf{a}) \,
\mathcal{N}(\mathbf{0},\mathbf{A},\mathbf{a}) \,
\nonumber \\
& & \mbox{} \times
p(\mathbf{\rho}|\tau,\theta,\phi,H_1) \,
\mathrm{d}\mathbf{a} \, \mathrm{d}\mathbf{\rho} \, \mathrm{d}\mathbf{h} \nonumber \\
&=&
\int_{V_{\rho}}
\mathcal{N}(\mathbf{0},(\mathbf{\Sigma}^{-1}-\mathbf{K})^{-1},\mathbf{x}) \,
p(\mathbf{\rho}|\tau,\theta,\phi,H_1) \,
\mathrm{d}\mathbf{\rho} \, ,
\end{eqnarray}
where the matrix
\begin{eqnarray}
\fl \mathbf{K}(\mathbf{\rho},\tau,\theta,\phi)&=&
(\mathbf{\Sigma}^{-1}\mathbf{F}\mathbf{W})
(
(\mathbf{F}\mathbf{W})^T
\mathbf{\Sigma}^{-1}
\mathbf{F}\mathbf{W}
+
\mathbf{A}^{-1}
)^{-1}
(\mathbf{\Sigma}^{-1}\mathbf{F}\mathbf{W})^T
\end{eqnarray}
will be the kernel of our numerical implementation.  Note that this is a generalization of
equation (\ref{eq:simpleC}) obtained in the single-sample case.  Since
\begin{eqnarray}\label{eqn:note}
\fl \frac{
        p(\mathbf{x}|\rho,\tau,\theta,\phi,H_1)
}{
        p(\mathbf{x}|H_0)
}
& = &
\frac{
        \mathcal{N}(\mathbf{0},(\mathbf{\Sigma}^{-1}-\mathbf{K})^{-1},\mathbf{x})
}{
        \mathcal{N}(\mathbf{0},\mathbf{\Sigma},\mathbf{x})
}
 =
\sqrt{|\mathbf{I}-\mathbf{\Sigma}\mathbf{K}|}
\exp(\frac{1}{2}\mathbf{x}^T\mathbf{K}\mathbf{x}) \, ,
\end{eqnarray}
we have
\begin{eqnarray}
\fl \frac{p(\mathbf{x}|\tau,\theta,\phi,H_1)}{
        p(\mathbf{x}|H_0)}
& = &
        \int_{V_{\rho}}
        p(\mathbf{\rho}|\tau,\theta,\phi,H_1)
        \sqrt{|\mathbf{I}-\mathbf{\Sigma}\mathbf{K}|}
        \exp(\frac{1}{2}\mathbf{x}^T\mathbf{K}\mathbf{x})
        \, \mathrm{d}\mathbf{\rho} \, . \label{eqn:explicit}
\end{eqnarray}
and the Bayes factor becomes
\begin{eqnarray}
\fl \frac{p(\mathbf{x}|H_1)}{p(\mathbf{x}|H_0)}
& = &
        \int_{V_{\rho, \tau, \theta, \phi}} \!\!\!\!\!\!
        p(\mathbf{\rho},\tau,\theta,\phi|H_1)
        \sqrt{|\mathbf{I}-\mathbf{\Sigma}\mathbf{K}|}
        \exp(\frac{1}{2}\mathbf{x}^T\mathbf{K}\mathbf{x})
        \, \mathrm{d}\mathbf{\rho} \, \mathrm{d}\tau
        \, \mathrm{d}\theta \, \mathrm{d}\phi \, .\label{eq:fastbayesfactor}
\end{eqnarray}
In other words we have reduced the task of computing the Bayes factor
to an integral over arrival time, source sky direction, and any additional signal
model parameters $\mathbf{\rho}$.

%
%

\subsection{Example signal models\label{sec:signalexamples}}

A simple signal model is the wideband signal model discussed briefly in Section~\ref{sec:wideband}.  This is a burst whose spectrum is white, has
characteristic strain amplitude $\sigma$ (at the Earth) and duration
$f_\textrm{s}^{-1}L$
\begin{eqnarray}
G&=&2L \label{wnb1}\\
\mathbf{A}&=&\sigma^2\mathbf{I}\label{eq:sigma} \label{wnb2} \\
\mathbf{W}&=&\mathbf{I} \, . \label{wnb3}
\end{eqnarray}

If we assert that such bursts are equally likely to come from any
source sky direction and arrive at any time in the observation window of
$f_\textrm{s}^{-1}M$ seconds, then the priors are
\begin{eqnarray}
p(\theta|H_1)&=&\frac{1}{2}\sin(\theta)\\
p(\phi|H_1)&=&(2\pi)^{-1}\\
p(\tau|H_1)&=&f_\textrm{s}M^{-1} \, .
\end{eqnarray}
If we assert that the source population is distributed uniformly in flat space up to some horizon $r_\mathrm{max}$
we have a prior on the distance $r$ to the source
$p(r|H_1)\propto r^2$.  We want to turn this into a prior on the characteristic amplitude $\sigma$, an example of a signal model
parameter we must numerically marginalize over
($\mathbf{\rho}=[\sigma]$).  Since the gravitational wave energy decays with the square of the distance to the source, $\sigma^2\propto r^{-2}$, we then deduce that:
\begin{eqnarray}
p(\sigma|H_1)&=&p(r|H_1)\left|\frac{\mathrm{d}r}{\mathrm{d}\sigma}\right|\\
&=&\frac{3\sigma_\mathrm{min}^3}{\sigma^4} \, ,\label{eq:sigma4prior}
\end{eqnarray}
where $\sigma_\mathrm{min} \propto r_\mathrm{max}^{-1}$ is a lower bound on the amplitude of (or upper bound on the distance of)
the gravitational wave.  This bound is obviously
somewhat arbitrary, but is a consequence of the way we distinguish
between detection and non-detection.  For a uniformly spatially
distributed population of bursts there are of course many weak signals
within the data, and the noise hypothesis is ``never'' true.  In reality
we are interested only in gravitational waves of at least a certain
size.  If $\sigma_\mathrm{min}$ is much smaller than the noise floor
in all detectors, the expression for the noise hypothesis is an
excellent approximation to the expressions of the likelihood we
adopted.  The classification of observations is insensitive to
different choices of $\sigma_\mathrm{min}$ below the noise floor.

This distribution of $\sigma$ is preserved if we consider a source population
with a distribution of different intrinsic luminosities, so long as they
are uniformly distributed in space out to their respective
$r_\mathrm{max}$ determined by the choice of $\sigma_\mathrm{min}$.

This is an example of a relatively \emph{uninformative} signal model.
It is capable of detecting signals of any waveform (of appropriate
duration).  However, it incurs a large {\it Occam penalty} for its
generality, and cannot be as sensitive as a more \emph{informed}
search.

The other extreme situation is where a source's waveform is completely
known, but its other parameters (amplitude, source sky position, polarization angle) are not.  Consider a source that
produces a linearly polarized strain $\mathbf{w}$.  If the source's
orientation, inclination and amplitude are unknown, we can
parameterize the system with two amplitudes $\mathbf{a}$ mapping the
strain into the observatory network's polarization basis
\begin{eqnarray}
\mathbf{W}&=&\left[
\begin{array}{cc}
\mathbf{w} & \mathbf{0}\\
\mathbf{0} & \mathbf{w}
\end{array}\right].
\end{eqnarray}
This is the Bayesian equivalent of the matched filter.
The template $\mathbf{w}$ appears twice because any specific signal
typically will not be aligned with the polarization basis used to describe
$h_+$ and $h_\times$ in the detectors, but rather will be rotated by
some {\em polarization angle} $\psi$ with respect to that basis.
More generally, any signal model that is independent of the observatory
network's polarization basis must have $\mathbf{A}$ and $\mathbf{W}$
composed of two identical sub-matrices on the diagonal like this, so that
$\mathbf{h}_+$ and $\mathbf{h}_\times$ have the same statistical distribution.  For
example, if the source is not linearly polarized, but has strain described
by $\mathbf{w}_+$ and $\mathbf{w}_\times$, then
\begin{eqnarray}
\mathbf{W}&=&\left[
\begin{array}{cccc}
\mathbf{w}_+ & \mathbf{w}_\times & \mathbf{0} & \mathbf{0}\\
\mathbf{0} & \mathbf{0} & \mathbf{w}_+ & \mathbf{w}_\times
\end{array}\right].
\end{eqnarray}

A more general case might be where we have a number of different
predictions for a waveform, $\mathbf{w}_i$, numerically derived. The
resulting search looks for a linear combination of these different
waveforms,
\begin{eqnarray}
\mathbf{W}&=&\left[
\begin{array}{cccccc}
\mathbf{w}_1 & \mathbf{w}_2 & \cdots & \mathbf{0} & \mathbf{0} & \cdots \\
\mathbf{0} & \mathbf{0} & \cdots & \mathbf{w}_1 & \mathbf{w}_2 & \cdots
\end{array}
\right] \, .
\end{eqnarray}

\subsection{Comparison with previously proposed methods}
\label{sec:comparison}

In this section we will expand on the arguments sketched in a previous
paper \cite{SeSuTiWo:08}.

%
%

Several previously proposed hypothesis tests, such as the
G\"{u}rsel-Tinto (i.e. standard likelihood), the constraint likelihoods,
and the Tikhonov-regularized likelihood, can be written in the form
\begin{eqnarray}\label{eq:prev}
        \max_{\rho,\tau,\theta,\phi}\mathbf{x}^T\mathbf{J}(\rho,\tau,\theta,\phi)\mathbf{x}&>&\lambda \, ,\label{eq:fht}
\end{eqnarray}
where $\mathbf{J}$ is an $MN\times MN$ matrix and $\lambda$ is a
\emph{threshold}.  These tests proceed in two steps.  First, parameters are
\emph{estimated} by maximizing the likelihood function with respect to the parameters.
Second, the value of the likelihood function at its maximum is compared to a threshold $\lambda$, which is chosen to ensure that it is only exceeded for the noise hypothesis at some acceptable \emph{false alarm rate}.

The corresponding Bayesian expression, from (\ref{eq:fastbayesfactor}),
integrates over source sky direction, arrival time and any other parameters
and determines if the Bayes factor is large enough to overcome the prior plausibility ratio
\begin{eqnarray}
\fl \int_{V_{\rho, \tau, \theta, \phi}} \!\!\!\!\!\!
        p(\mathbf{\rho},\tau,\theta,\phi|H_1)
        \sqrt{|\mathbf{I}-\mathbf{\Sigma}\mathbf{K}|}
        \exp(\frac{1}{2}\mathbf{x}^T\mathbf{K}\mathbf{x})
        \, \mathrm{d}\mathbf{\rho} \, \mathrm{d}\tau
        \, \mathrm{d}\theta \, \mathrm{d}\phi
        &>&
\frac{
p(H_0)
}{
p(H_1)
}\,. \label{eq:bht}
\end{eqnarray}

There are some obvious similarities between (\ref{eq:fht}) and (\ref{eq:bht}),
in particular the quadratic forms central to each.  However, direct mathematical equivalence cannot be established in
general because of the difference between maximization and marginalization.

We can establish equivalence for the related problem of parameter estimation, where we have maximum likelihood parameter estimate
\begin{eqnarray}
\{\rho,\tau,\theta,\phi\}&=&\arg\max(\mathbf{x}^T\mathbf{J}\mathbf{x})
\end{eqnarray}
and the Bayesian most plausible parameters, one of several ways the posterior plausibility distribution for the parameters can be turned into a point estimate
\begin{eqnarray}
\fl \{\rho,\tau,\theta,\phi\}&=&\arg\max ( p(\mathbf{\rho},\tau,\theta,\phi|H_1)
        \sqrt{|\mathbf{I}-\mathbf{\Sigma}\mathbf{K}|}
        \exp(\frac{1}{2}\mathbf{x}^T\mathbf{K}\mathbf{x}) )\\
        &=&\arg\max(\mathbf{x}^T\mathbf{K}\mathbf{x} + 2\ln p(\mathbf{\rho},\tau,\theta,\phi|H_1) + \ln |\mathbf{I}-\mathbf{\Sigma}\mathbf{K}|) \, .
\end{eqnarray}
In the cases where we can find a Bayesian signal model that produces $\mathbf{K}=\mathbf{J}$, we must also use a prior
\begin{eqnarray}
p(\mathbf{\rho},\tau,\theta,\phi|H_1)&\propto&|\mathbf{I}-\mathbf{\Sigma}\mathbf{K}|^{-\frac{1}{2}}.
\end{eqnarray}
This prior states that gravitational wave bursts are \emph{intrinsically} more likely to occur at the sky positions
that the network is more sensitive to.  We interpret this as an implicit bias present in any statistic of the form of (\ref{eq:fht})\footnote{It is important
to note that this particular objection applies only to all-sky searches;
it is a consequence of the maximization over $(\theta,\phi)$.  These statistics
are also used in directed searches (for example, in the direction of a gamma-ray burst) where
$(\theta, \phi)$ is known and fixed, and the problem does not arise (the missing normalization
term is one of several absorbed by tuning the threshold).}.

%

In order to compare previously proposed statistics to the Bayesian
method, we place some restrictions on the configurations
considered.  We will assume co-located (but differently oriented)
detectors to eliminate the need to time-shift data, and we will use
stationary signals and observation times that coincide with the time
the signal is present.  These restrictions eliminate the differences
in the way previously proposed statistics and the Bayesian method
handle arrival time and signal duration.  For simplicity, we will
further assume that the detectors are affected by white Gaussian noise.
The conclusions drawn will apply equally to different versions of
these statistics for colored noise or different bases other than the
time-domain (such as the frequency or wavelet domains).

\subsubsection{Tikhonov regularized statistic}

The Tikhonov regularized statistic proposed in \cite{Ra:06} for white
noise interferometers is
\begin{eqnarray}
\mathbf{x}^T\mathbf{F}(\mathbf{F}^T\mathbf{F}
+\alpha^2\mathbf{I})^{-1}\mathbf{F}^T\mathbf{x}\, .
\end{eqnarray}
The Bayesian kernel $\mathbf{K}$ reduces to this for
\begin{eqnarray}
\mathbf{\Sigma}&=&\mathbf{I}\\
\mathbf{W}&=&\mathbf{I}\\
\mathbf{A}&=&\alpha^{-2}\mathbf{I} \, .
\end{eqnarray}
This is a signal of
characteristic amplitude $\sigma = \alpha^{-1}$.  The Tikhonov
regularizer $\alpha$ therefore places a delta function prior on the characteristic amplitude of the signal $p(\sigma|H_1)=\delta(\sigma-\alpha^{-1})$.

The Tikhonov statistic behaves like a Bayesian statistic that
postulates all bursts have energies in a narrow range. 

\subsubsection{G\"{u}rsel-Tinto statistic}

The G\"{u}rsel-Tinto or standard likelihood statistic
\cite{GuTi:89,FlHu:98b,AnBrCrFl:01} is
\begin{eqnarray}
\mathbf{x}^T\mathbf{F}(\mathbf{F}^T\mathbf{F})^{-1}\mathbf{F}^T\mathbf{x}\,.
\end{eqnarray}
For large $\sigma$, the Tikhonov statistic goes to
\begin{eqnarray}
\mathbf{K}
  &\approx&  \mathbf{F}(\mathbf{F}^T\mathbf{F})^{-1}\mathbf{F}^T \, .
\end{eqnarray}
This implies that the G\"{u}rsel-Tinto statistic is the limit of a series of Bayesian statistics for increasing signal amplitudes.

\subsubsection{Soft constraint likelihood}

The soft constraint statistic \cite{KlMoRaMi:05,KlMoRaMi:06} for white
noise interferometers is
\begin{eqnarray}\label{eqn:SC}
k^2(\theta,\phi)\,\mathbf{x}^T\mathbf{FF}^T\mathbf{x} \, ,
\end{eqnarray}
for some function $k(\theta,\phi)$.  Specifically, (\ref{eqn:SC})
gives the soft constraint likelihood for the choice
$k^2=(\mathbf{F}^{+T}\mathbf{F}^+)^{-1}$, where the antenna response is
computed in the dominant polarization frame \cite{KlMoRaMi:05}.

Consider the signal model defined by
\begin{eqnarray}
\mathbf{\Sigma}&=&\mathbf{I}\\
\mathbf{W}&=&\mathbf{I}\\
\mathbf{A}&=&\sigma^2k^2(\theta,\phi)\mathbf{I} \, .
\end{eqnarray}
This is a population of signals whose characteristic amplitude
$\sigma k(\theta,\phi)$ varies as some known function of source sky direction,
slightly generalizing the situation of the Tikhonov statistic.  For small $\sigma$,
\begin{eqnarray}
\mathbf{K}&\approx&\sigma^2k^2(\theta,\phi)\mathbf{F}\mathbf{F}^T \, ,
\end{eqnarray}
so we can see that the soft constraint is the limit of a series of Bayesian statistics for decreasing signal amplitudes.

\subsubsection{Hard constraint likelihood}

Let us restrict the soft-constraint signal model to a population of
\emph{linearly polarized} signals with a known polarization angle
$\psi(\theta,\phi)$ for each source sky direction
\begin{eqnarray}
\mathbf{\Sigma}&=&\mathbf{I}\\
\mathbf{W}&=&\left[
\begin{array}{c}
\cos 2\psi(\theta,\phi)\mathbf{I}\\
\sin 2\psi(\theta,\phi)\mathbf{I}
\end{array}
\right]\\
\mathbf{A}&=&\sigma^2 k^2(\theta,\phi)\mathbf{I} \, .
\end{eqnarray}
Then 
for $\sigma\rightarrow 0$ the Bayesian statistic limits to
\begin{eqnarray}
k^2(\theta,\phi) \, \mathbf{x}^T\mathbf{FW}(\mathbf{FW})^T\mathbf{x}
\, .
\end{eqnarray}
For the particular choice of $\psi(\theta,\phi)$ being the rotation
angle between the detector polarization basis and the dominant
polarization frame, and $k^2=(\mathbf{FW})^T\mathbf{FW}$ (which is
equal to ($\mathbf{F}^{+T}\mathbf{F}^+)^{-1}$ in the dominant
polarization frame \cite{KlMoRaMi:05}), this yields the hard
constraint statistic of \cite{KlMoRaMi:05}.

In addition to the explicit assumptions that all signals are
linearly polarized with known polarization angle, the hard constraint
has the same properties as the soft constraint.

\subsection{Interpretation}

We have shown that several previously proposed statistics are special cases
or limiting cases of Bayesian statistics for particular choices of prior.
%
The `priors' implicit in these non-Bayesian methods are not representative of our
expectations about the source population, so we can reasonably expect
improved performance from a detection statistic with priors
better reflecting our state of knowledge.  The Bayesian analysis allows us to begin with our physical understanding
of the problem, described in terms of prior expectations about the
gravitational wave signal population, and derive the detection
statistic for these conditions.  The effects of priors are lessened when there is a strong gravitational wave
signal present; all these statistics, Bayesian and non-Bayesian, are effective at
detecting stronger gravitational waves; significant differences occur only
for marginal signals.  In the next section, we will quantitatively compare the relative performance
of the methods mentioned above and the Bayesian statistic we propose.

%% file: results.tex
\section{Simulations \label{sec:simulations}}
\label{SECIII}

To
characterize the relative performance of the G\"{u}rsel-Tinto (i.e.
standard likelihood), soft constraint, hard constraint, and Bayesian
methods we used the \textsc{X-Pipeline} software package \cite{Xpipeline}.
This package reads in gravitational wave data, estimates the power spectrum and whitens the data, and transforms it into a time-frequency basis of successive short Fourier transforms.  Each statistic is then applied to the transformed data, and the results saved to file.  This ensures that the observed differences are due to the statistics themselves, and not to different whitening or other conventions.

Our tests used a set of 4 identical detectors at the positions and
orientations of the LIGO-Hanford, LIGO-Livingston, GEO 600, and Virgo
detectors.  The data was simulated as Gaussian noise with spectrum
following the design sensitivity curve of the 4-km LIGO detectors; it
was taken from a standard archive of simulated data \cite{Be_etal:05}
used for testing detection algorithms.  Approximately 12 hours of data
in total was analysed for these tests.

\begin{figure}
%
\includegraphics[width=\textwidth]{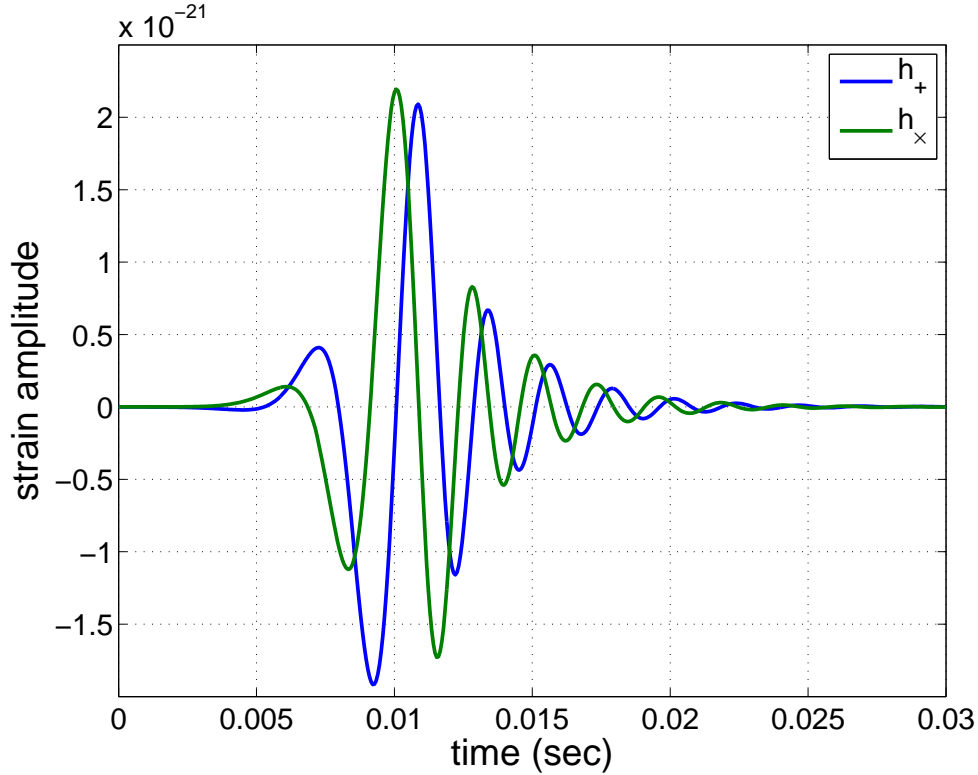}
\caption{\label{fig:lazarus-timeseries}
  Time-series Lazarus waveforms \cite{Ba_etal:02} used for our
  simulations, from a nominal distance of 240 Mpc.  }
\end{figure}

\begin{figure}
\includegraphics[width=\textwidth]{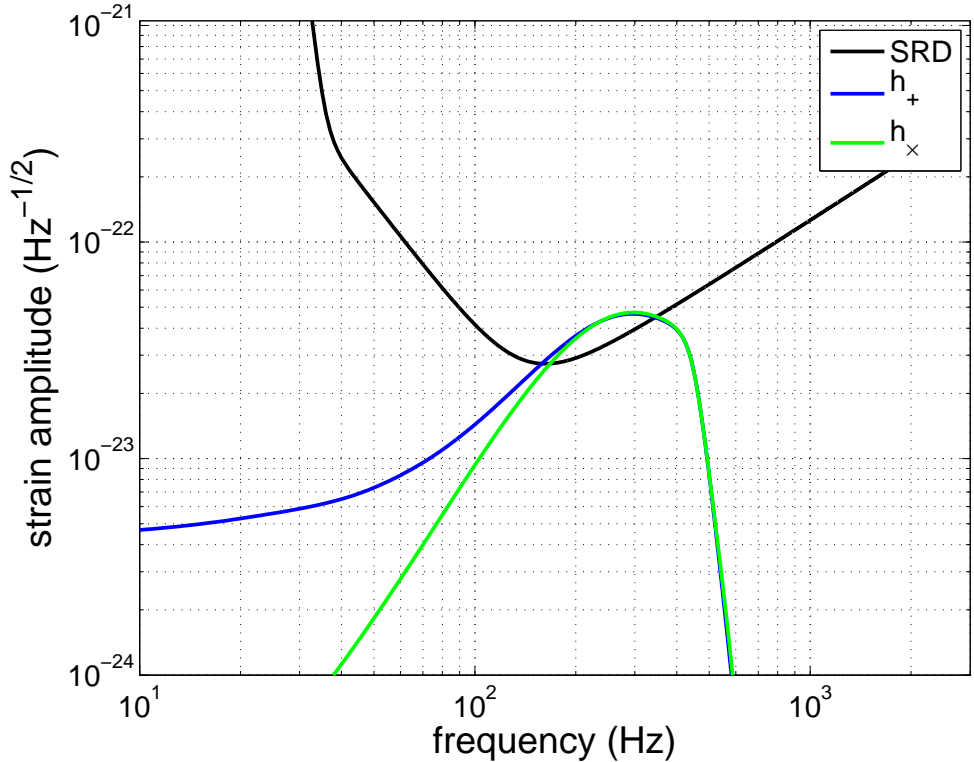}
\caption{\label{fig:lazarus-freqseries}
  Strain-equivalent noise amplitude spectral density of the simulated
  data used in our tests (black) with spectra of the Lazarus waveforms
  \cite{Ba_etal:02} at 240 Mpc.  The Lazarus spectra have been rescaled
  by $T^{-1/2}=(128/\mathrm{sec})^{1/2}$ to render them into the same units
  as the noise spectrum.}
  \end{figure}

For the population of gravitational-wave signals to be detected we
chose, somewhat arbitrarily, the ``Lazarus'' waveforms of Baker {\em
  et al.\/} \cite{Ba_etal:02}.  These are fairly simple waveforms
generated from numerical simulations of the merger and ringdown of a
binary black-hole system.  We chose to simulate a pair of 20
solar-mass black holes, which puts the peak of signal power near the
frequencies of best sensitivity for LIGO.  The time-series waveforms
are shown in Fig.~\ref{fig:lazarus-timeseries}, while the spectra and
detector noise curve are given in Fig.~\ref{fig:lazarus-freqseries}.
The sources were placed at the discrete distances $240/S$ Mpc
\footnote{The fiducial distance $240$ Mpc is chosen for numerical
convenience; it is the distance at which the sum-squared matched-filter
SNR for each polarization is $1/2$, assuming optimal antenna response
($F^+,F^\times=1$).},
where $S = 1, 2, 2.5, 3, 10$, and with randomly chosen sky position and
orientation. Approximately 5000 injections were performed for each
distance.

%
%

For the Bayesian statistic we adapt the broad-band signal prior
(\ref{wnb1})--(\ref{wnb3}) for each polarization.  Specifically, we assume the
simple model of a burst whose spectrum is white, with characteristic
strain amplitude $\sigma$ at the Earth and duration equal to our chosen
FFT length:
\begin{eqnarray}
G &=&  2M\\
\mathbf{A}  &=&  \sigma^2\mathbf{I}\label{eq:sigma}\\
\mathbf{W}  &=&  \mathbf{I} \, .
\end{eqnarray}
We use a uniform prior on the signal arrival time $\tau$, and an isotropic
prior on the sky position $(\theta,\phi)$.
The Bayesian statistic was computed for approximately logarithmically
spaced discrete values of characteristic strain 
$\sigma = 10^{-23}, 3\times10^{-23}, 10^{-22}, 3\times10^{-22}, 10^{-21}$,
and averaged together in post-processing.  This averaging approximates
a single Bayesian statistic with a Jeffreys (scale invariant) prior
$p(\sigma)\propto1/\sigma$ between $10^{-23}$ and $10^{-21}$.
(Performing the combination in post-processing allowed us to maintain
compatibility with the existing architecture of \textsc{X-Pipeline}; we do not use the $\sigma^{-4}$ prior from (\ref{eq:sigma4prior}) because we are injecting from a fixed distance, not a spatially uniform population.)

\begin{figure}
\includegraphics[width=\textwidth]{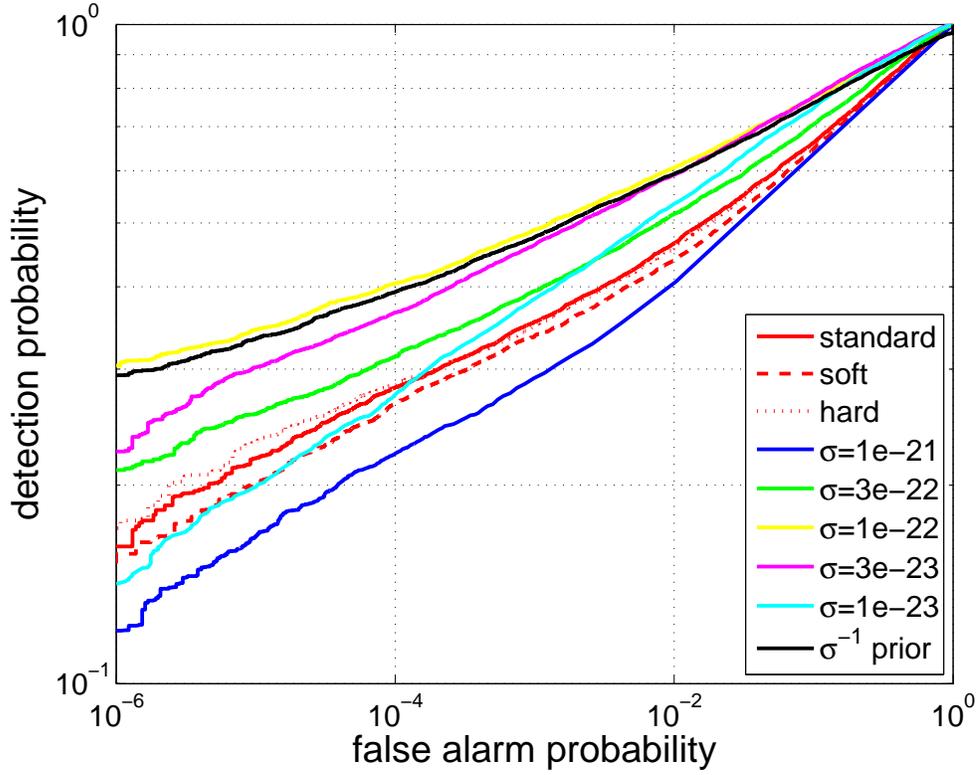}
\caption{Receiver-operating characteristic (ROC)
  curves for the Bayesian, standard (G\"ursel-Tinto), soft
  constraint, and hard constraint likelihoods for sources at 96 Mpc.  The curve for a $\sigma^{-1}$
  prior is obtained by marginalizing over probabilities
  associated with the discrete $\sigma$ values tested.  The best
  performance is achieved by the $\sigma$ value most closely matching
  the amplitude of the injected signals, with the marginalized curve performing almost
  identically.  The detection probability of the marginalized Bayesian
  method is significantly greater than that of any of the non-Bayesian
  methods (standard/G\"ursel-Tinto, soft constraint, and hard
  constraint likelihoods) over the full range of false-alarm
  probabilities tested.}
\label{fig:ROC}
\end{figure}

Each likelihood statistic was computed over a fixed frequency band of
[64,1088] Hz, with an FFT length of 1/128 sec.  We analyzed blocks of data, overlapping by 75\% of their duration.  The detection probability as a function of
false alarm probability is shown in Fig.~\ref{fig:ROC}.  The distance
used for the Lazarus simulations for this figure was
$240/2.5=96$~Mpc; injections at other distances yielded similar
results.  At this distance, the total SNR deposited in the network
$\sqrt{\sum_\alpha \mathrm{SNR}^2}$ was in the range $\sim1-8$ with a mean value of
5, where
\begin{equation}
\sum_\alpha \mathrm{SNR}^2_\alpha
  =  \sum_\alpha 4 \int_0^\infty \!\! df \, \frac{\left|
          F_\alpha^+ \tilde{h}_+(f)
          + F_\alpha^\times \tilde{h}_\times(f)
      \right|^2}{S(f)}
\end{equation}
and $S(f)$ is the one-sided noise power spectral density of each interferometer.

Fig.~\ref{fig:ROC} is the receiver-operating characteristic plot for
each of the statistics considered.  The vertical axis represents the
fraction of a population of signals whose detection statistics exceed the
threshold that would only be crossed by background noise at the rate
given by the false alarm probability on the horizontal axis.
For example, we can read off the figure that if we can afford a false alarm probability of $10^{-2}$, the various detection
statistics are able to detect between $0.4$ and $0.6$ of the injected signals.
We see that the best performance
is
achieved by the Bayesian method with the $\sigma$ value most closely
matching the injected signals, with the marginalized curve performing
almost identically.  The detection probability of the marginalized
Bayesian method is significantly better than that of any of the
non-Bayesian methods (standard likelihood, soft constraint, and hard
constraint likelihoods) over the full range of false-alarm
probabilities tested.

For a given false-alarm probability, we may compute the
distance at which each likelihood achieves 50\% efficiency by fitting
a sigmoid curve to the simulations.  The observed volume, and
therefore the expected rate of detections for a uniformly distributed
source population, scales as the cube of the distance.  We computed
the distance and volume for each statistic for two false alarm probabilities,
$10^{-5}$ in Table~\ref{table:1e-5} and $1/256\approx 3.9\times
10^{-3}$ in Table~\ref{table:1-256}.  As we compute 512 statistics per
second, these correspond to false alarm rates of 1/200 Hz and 2 Hz
respectively (as the statistics are computed on 75\% overlapped data,
these estimates are conservative).
These rates are practical for event generation
at the first stage of an untriggered (all-sky, all-time) burst search \cite{Ab_etal:05c,Ab_etal:08,Ab_etal:07rh,Ab_etal:07}.
At both of these false alarm probabilities, the Bayesian
method can detect sources approximately 15\% more distant, and
consequently has an observed volume and expected detection rate
approximately 50\% greater, than the non-Bayesian statistics.

It is important to note that we do {\em not} use detailed knowledge of
the signal waveform for the Bayesian analysis.  Our prior is that the signal
spectrum is flat over the analysis band ([64,1088] Hz), and by imposing
no phase structure or sample-to-sample correlations we are assuming that
over the integration time (1/128 sec) the time samples of strain are independently and identically distributed.
Considering Figures~\ref{fig:lazarus-timeseries} and~\ref{fig:lazarus-freqseries},
it is clear that these priors are not particularly accurate models for the
actual gravitational-wave signal.  Nevertheless, our Monte Carlo results
demonstrate that even this incomplete prior information can improve the sensitivity of the search.

\begin{table}
\caption{Distances for false-alarm probability $10^{-5}$}
\begin{tabular}{cccc}
\hline\hline
Statistic & Distance (Mpc) & Distance (rel.) & Volume (rel.) \\
\hline
Standard & 72.1 & 1.01 & 1.02 \\
Soft & 71.6 & 1.00 & 1.00 \\
Hard & 72.5 & 1.01 & 1.04 \\
Bayesian & 82.0 & 1.15 & 1.50 \\
\hline
\end{tabular}
\label{table:1e-5}
\end{table}

\begin{table}
\caption{Distances for false-alarm probability $1/256$}
\begin{tabular}{cccc}
\hline\hline
Statistic & Distance (Mpc) & Distance (rel.) & Volume (rel.) \\
\hline
Standard & 87.1 & 1.03 & 1.08 \\
Soft & 84.9 & 1.00 & 1.00 \\
Hard & 86.9 & 1.02 & 1.07 \\
Bayesian & 97.9 & 1.15 & 1.53 \\
\hline
\end{tabular}
\label{table:1-256}
\end{table}


%% file: conclusions.tex


\section{Conclusions and Future directions}
\label{SECIV}

We have presented a comprehensive Bayesian formulation of the
problem of detecting gravitational wave bursts with a network of
ground-based interferometers.  We demonstrated how to
systematically incorporate prior information
into the analysis, such as time-frequency or polarization content,
source distributions and signal strengths.  We have also seen that this Bayesian
formulation contains several previously proposed detection statistics
as special cases.

The Bayesian methodology we have derived to address the problem of
detecting poorly-understood gravitational wave bursts yields a novel
statistic.  On theoretical grounds we expect this statistic to
outperform previously proposed statistics.  A Monte-Carlo analysis
confirms this expectation: over a range of false alarms rates, the
Bayesian statistic can detect sources at 15\% greater distances and
therefore observe 50\% more events.

The Bayesian search requires explicitly adopting a model for the
poorly understood signal.  This is not a shortcoming.  The model may
be agnostic with respect to many features of the waveform.  As we
have demonstrated, existing methods are not free from their own
signal models, but implicitly assume priors on the energies and spatial distribution of sources.

Coherent analyses of any kind are relatively costly, and efficient
implementations must be sought.  By specifying the problem as an
integral, the Bayesian approach lets us leverage the extensive
literature on numerical integration for techniques to accelerate the
computation; one promising contender is \emph{importance sampling}.

As second practical issue that must be dealt with is that the background
noise of real gravitational wave detectors contains transient non-Gaussian
features (``glitches'').  As in the case of detection statistics, several
{\em ad hoc} non-Bayesian statistics have been proposed to distinguish
glitches from gravitational wave signals (see for example \cite{Ch_etal:06,WeSc:05}).
Again, Bayesian methodology provides us with a direction in which to
proceed: augment the noise model to better reflect ``glitchy'' reality,
and to the extent we are successful, robustness will automatically follow.
